# A Learning Model Applied to the Calculation of the Velocities of 20 Stars Relative to the Sun

Rafael Edgardo Carlos Reyes[*1,2], Atilio Buendía Giribaldi[2], Felipe Américo Reyes Navarro[1]

[1]Facultad de Ciencias Físicas, Universidad Nacional Mayor de San Marcos, Lima, Perú
[2]Universidad Interamericana para el Desarrollo, Lima, Perú

**Abstract**

We aim to explain the paradigm of a learning model, as well as to validate it in an applied case of an astronomy problem where the data used are declination, parallax, radial velocity of a star, as well as its annual variation in right ascension and declination. This study is based on a socio-critical and positivist paradigm in the context of basic and applied science; algorithms and astronomical models were used as an instrument, which allowed us to address a specific case such as the calculation of the velocity of a star relative to the Sun.

## Um Modelo de Aprendizagem Aplicado ao Cálculo da Velocidades de 20 Estrelas relativas ao Sol

**Resumo**

O objetivo desta pesquisa é explicar o paradigma de um modelo de aprendizagem, bem como validá-lo em um caso aplicado de um problema de astronomia onde os dados utilizados são a declinação, paralaxe, velocidade radial de uma estrela, bem como sua variação em ascensão reta e declinação. Este estudo assenta num paradigma sócio crítico e positivista num contexto de ciência básica e aplicada, tendo como instrumentos utilizados os algoritmos e modelos astronômicos que nos permitiram abordar o caso concreto do cálculo da velocidade de uma estrela relativa ao Sol.

**Palavras-chave:** Modelo de Aprendizagem, conectivismo, velocidade de uma estrela em relação ao sol.

# 1. Introduction

The most studied theories of learning—behaviorism, cognitivism, constructivism, and social learning theory, which seek to understand how the person learns—are based on epistemological traditions such as objectivism, pragmatism, and interpretivism. Nowadays, these theories are revised due to the advancement of technology. Until not long ago, people went to an educational institution for technical or professional training; a time of around a decade was required for this. Currently, knowledge is increasing exponentially in almost all areas. We can cite, for example, the case of astronomy, where the information collected by different space satellites already exceeds thousands of terabytes. Not all this amount of data is processed or analyzed, while waiting for tools that allow working with this enormous amount of information.

Among the limitations of the above theories, we find they consider that learning occurs within a person. They are not concerned with learning occurring outside of a person or within organizations [1] nor are they concerned with the value of what is being learned (nonetheless, we must mention organizational learning [2]). Then, before explaining the learning model used in this article, we will now explain the situation prior to its appearance.

During the 1970s, some scientists from the US and Europe—such as David Ruelle, Robert May, and Edward Feigenbaum—developed chaos theory to study irregular, discontinuous, and erratic behavior in nature, such as disorder in the atmosphere, turbulence at sea, fluctuations in wildlife populations, oscillations of the heart and brain [3]. Today, chaos theory is investigated in areas such as mathematics, physics, biology, meteorology, economics, etc., that is, the use of chaos theory is multidisciplinary. In the context of this theory, the word "chaos" refers to a behavior that is both highly sensitive to the initial conditions and presents the property of recurrence [4]. This type of behavior can be found in deterministic systems like some nonlinear dynamical systems, but it can also be present in complex systems, which are not deterministic systems. Historically, it was even called deterministic chaos to differentiate it from the chaos created by chance. This made sense, since in chaos theory there are well-defined formulas and rules for studying a certain dynamical system.

The development of current technology has allowed the appearance of networks, which are connections between entities, as well as nodes competing for connections between those entities because the links represent survival in an interconnected world [5]. The central premise is that unusual node connections support and intensify large effort activities.

Technology is helping to store and process a lot of cognitive information. Social networks have become an additional element in learning during the digital age. Additionally, Karen Stephenson's concept *the quantum theory of trust* explains not only how to realize the collective cognitive capacity of an organization but also how to fertilize and increase it [6]. The interconnected hubs can promote and maintaining the effective flow of knowledge; they are very important for the activities in an organization.

From what was mentioned above, the learning model known as connectivism emerged almost two decades ago [1,7]. According to Siemens (2005) *"This amplification of learning, knowledge, and understanding through the extension of a personal network is the epitome of connectivism"* [1, 8]. We must stress that connectivism is not a finished theory. It is a research field. It seeks the integration of the principles explored by the theories of chaos, network, complexity, and self-organization, providing information on learning abilities and the tasks necessary to thrive in the digital age [9].

Within connectivism, learning can be a process that occurs with agents or environments varying over time, which are not necessarily under the control of the person. Therefore, it is essential to know what new information is acquired on a continuous way, as well as to distinguish between unnecessary and necessary information [1]. Connectivism also address the challenges of knowledge management within corporations, where creating, preserving, and using the flow of information must be a key organizational activity.

For connectivism, learning begins with the individual, while knowledge is constituted by a network feeding organizations and institutions, the same ones that feed back the network providing learning to the individual, forming a cycle: person → network → organization. This to facilitate continuous learning, keeping it current and updated through the connections formed by it, where the most important are those that allow to learn more. Learning and work are no longer separate activities, but often occur at the same time [1].

New information can alter the outlook, thus affecting a decision made earlier. An answer may be correct now, but incorrect tomorrow, due to the change in information. Therefore, following Siemens (2005) [1]:

*"The ability to synthesize and recognize connections and patterns is a valuable skill…As knowledge continues to grow and evolve, access to what is needed is more important than what the learner currently possesses."*

This becomes a vital skill. The *know-how* is being replaced by the *know-where*, that is, where to find the necessary knowledge.

## 2. Theoretical framework

### 2.1 Exponential Pedagogical Model

A new learning model is proposed when there is a need to know something, or it is also the product of some unsought random event. From Jean Piaget's constructivism through Kurt Lewin's action research methodology, we arrive at George Siemens' connectivism, applying open science and chaos theory. According to UNESCO 2022, open science is a movement that seeks to make science more accessible, efficient, transparent, and beneficial to all of us [10].

All future professionals must have a humanistic training, which undoubtedly also includes science to provide them a comprehensive image of the Universe surrounding us. Something that attracts the students, that questions their most common ideas and that can be dealt with some analysis, reflection and, even more, exercises.

Professorial work should not be limited to the transmission of knowledge, because we get mere retainers of knowledge [11] or, in the best case, inquisitive beings, dedicated to seeking and discovering knowledge, but with a poor spiritual life and a limited vision of things, even ignorant of the realities of the country itself, oblivious to the environment and its cultural and human wealth. The student must know how to classify the data, and update himself continuously, which improves the skills of his specialty. Many times, the academic information received lasts a short time, as is the case with the software packages that are updated with each new version

By considering these premises, we seek to embody our model by choosing an astronomy topic that allows the students not only to become familiar with the cosmos but also to apply philosophy and logic, developing its algorithm and a program in Python language allowing them to solve a problem in a systematic way. This can go on improving and growing to the extent of the needs that their information increases

**2.2 Learning Paradigm in an Exponential Context**

This consists of a proposal with the following sequence, the same one that we will address independently for the problem chosen, except for artificial intelligence that due to the relative simplicity of the problem has not been necessary. The sequence is the following:

| Philosophy → Computational Logic → Algorithm → Language Programming → Artificial Intelligence |
|---|

*2.2.1 Philosophy*

We begin first by remembering the famous Greek philosopher Pythagoras, who had a great capacity for observation and educated in an open way. This way he amalgamated theory and practice; as his knowledge increased, his curiosity also grew [12]. Among the sciences, astronomy is mainly observational, since it is not feasible to replicate in a laboratory everything observed in the Universe. For this reason, we will concentrate on a case that involves data obtained by observation and, by applying our knowledge of physics, we will obtain information regarding the relative speed of the stars.

The stars do not remain fixed in the sky; taking away the movement of the earth's rotation, careful observation shows that the relative positions between them change over time. The further away the stars are, the less perceptible this movement is. Thus, we can measure the relative movement of the closest stars with respect to the very distant ones, because the latter practically do not move.

By using an angular coordinate system such as the absolute equatorial system, which uses right ascension and declination angles, we can measure the change in these after one year. These changes will be called components of the proper motion of a star. Now, if we know not only the distance to the star (i.e., its parallax) but also its radial velocity, then it is possible to calculate its velocity relative to the sun.

## 3. Methodology

This research is based on a socio-critical and positivist paradigm in the context of basic research. The method consists in choosing a problem of interest that we wish to solve. First, we discuss the philosophy of the chosen theme, which consists of a set of reflections on the essence, properties, causes and effects of natural things. Second, we develop a logical structure of the problem to be solved, this will allow the development of a method of reasoning, providing rules and techniques to determine whether a given argument is valid. Third, now we elaborate an algorithm in such a way that it orders the systematic operations that allow to make a calculation and find the solution to the problem chosen. Fourth, we developed a program in the Python programming language, where the logical structure designed is maintained, as well as the algorithm developed for the problem. The fifth and last step corresponding to artificial intelligence is optional and is reserved for special problems that require machine learning and automation. Although for simpler problems it is possible to project its use as an extension to a recurring solution, but with updated data. In our example chosen, using artificial intelligence, large amounts of data can be combined with fast, iterative processing and smart algorithms, allowing the software to automatically learn to recognize patterns or features in the data, which can be continually updated, enabling even be used in space flight navigation.

## 4. Results and discussions

### 4.1. Mathematical and computational logic

Now we are going to describe the logical mathematical foundation that qualifies the solution of the problem [13].

Input data:
- declination $\delta$
- variation in right ascension $\mu_\alpha$
- Variation in declination $\mu_\delta$
- parallax $P$
- Radial velocity $v$

To build a logical structure of the problem, we are going to define some quantities such as:

P: Horizontal component of positive proper motion
Q: Vertical component of positive proper motion
~P: Horizontal component of negative proper motion
~Q: Vertical component of negative proper motion
R1: The position angle of the proper motion is in the first quadrant
R2: The position angle of the proper motion is in the second quadrant
R3: The position angle of the proper motion is in the third quadrant
R4: The position angle of the proper motion is in the fourth quadrant

To determine the position angle of the proper motion of a star, one of the following relationships is correct:

If (P ∧ Q) ⇒ R1
If (P ∧ ~Q) ⇒ R2
If (~P ∧ ~Q) ⇒ R3
If (~P ∧ Q) ⇒ R4

Thus, we obtain the position angle of the proper motion of the star,

T={[(P ∧ Q) ⇒ R1] V [(P ∧ ~Q) ⇒ R2] V [(~P ∧ ~Q) ⇒ R3] V [(~P ∧ Q) ⇒ R4]}

Knowing the angle, we can obtain the proper motion U, and knowing the parallax of the star, we can calculate its transverse speed W.

Knowing the components of the velocity, the position angle of the velocity is determined. For this we define the following quantities:

A: Positive transverse component of velocity
B: Positive radial component of velocity
~B: Negative radial component of velocity
S1: The position angle of velocity is in the first quadrant
S2: The position angle of velocity is in the second quadrant

To determine the position angle of the velocity of the star, one of the following relationships is correct:

If (A) ∧ B) ⇒ S1
If (A ∧ ~B) ⇒ S2,

this way we get the angle of position or direction of the speed of the star:
M={[(A ∧ B) ⇒ S1] V [(A ∧ ~B) ⇒ S2]} .

Finally, we can calculate the velocity relative to the sun, N:

{[(T ⇒ U) ⇒ W] ⇒ M} ⇒ N

## 4.2 Algorithm

By using the result of the logical-mathematical interpretation of the problem, we elaborated an algorithm (Figure 1), which summarizes the procedure to follow for the solution of the chosen problem [14, 15].

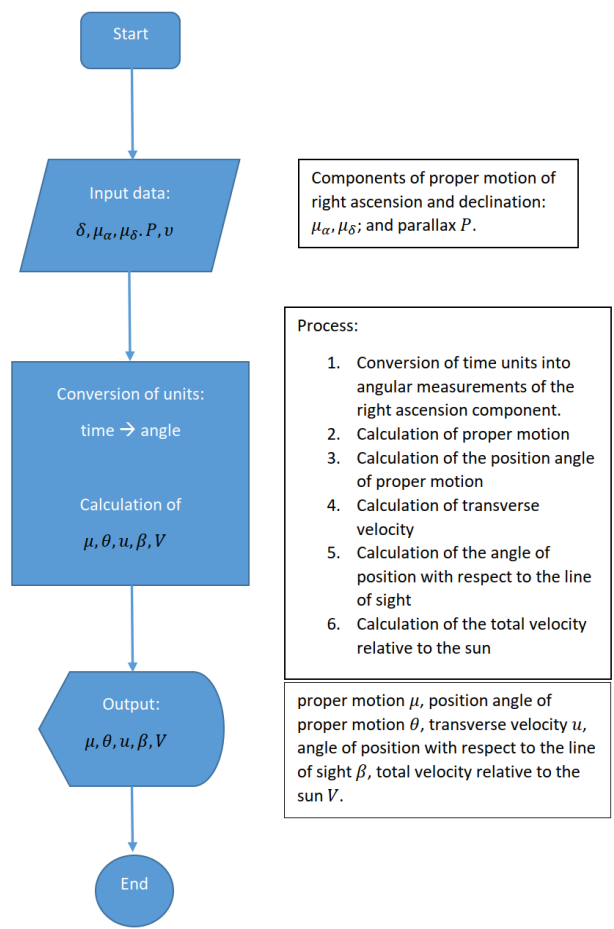

**Figure 1.** Algorithm summarizing the procedure for calculating the velocity of a star relative to the sun and other parameters too.

### 4.3 Data obtained

Table 1 lists the data used corresponding to 20 stars close to Earth, where the first column indicates enumeration; the second column, the star name; the following columns are the star data such as declination, right ascension variation, declination variation, parallax, and radial velocity. Besides, in Table 2, we have the results obtained after the corresponding process.

**Table 1.** Data used corresponding to 20 stars close to Earth.

| No. | Star name | Declination $\delta(°)$ | Right ascension variation $\mu_\alpha$ ("/year) | Declination variation $\mu_\delta$ ("/year) | Parallax (") | Radial velocity (km/s) |
|---|---|---|---|---|---|---|
| 1 | Barnard's Star | +04.69 | -0.802 | +10.362 | 0.547 | -110.1 |
| 2 | Kapteyn's Star | -45.02 | +6.491 | -5.709 | 0.254 | +245.1 |
| 3 | Groombridge 1830 | +37.72 | +4.003 | -5.818 | 0.109 | -98.0 |
| 4 | Lacaille 9352 | -35.85 | +6.766 | +1.330 | 0.304 | +9.7 |
| 5 | Ross 619 | +08.77 | +1.069 | -5.094 | 0.148 | +14.6 |
| 6 | 61 CYG A | +38.74 | +4.164 | +3.249 | 0.286 | -64.4 |
| 7 | Lalande 21185 | +35.97 | -0.580 | -4.766 | 0.393 | -85.6 |
| 8 | WX UMA | +43.52 | -4.339 | +0.961 | 0.204 | +70.0 |
| 9 | Wolf 489 | +03.68 | -3.736 | -1.114 | 0.119 | +183.4 |
| 10 | Arcturus | +19.18 | -1.094 | -1.999 | 0.089 | -5.2 |
| 11 | Proxima Centauri | -62.68 | +3.775 | +0.769 | 0.769 | +21.7 |
| 12 | BD +5 1668 | +05.23 | +0.571 | -3.694 | 0.264 | +18.4 |
| 13 | mu CAS | +54.92 | +3.422 | -1.599 | 0.130 | -98.3 |
| 14 | alpha Centauri A | -60.83 | -3.679 | +0.474 | 0.751 | -21.4 |
| 15 | Lacaille 8760 | -38.87 | -3.259 | -1.145 | 0.252 | +20.7 |
| 16 | Luyten 789-6 | -15.30 | +2.314 | +2.295 | 0.294 | -59.9 |
| 17 | Ross 451 | +67.26 | +0.261 | -3.159 | 0.040 | -118.0 |
| 18 | 82 ERI | -43.07 | +3.038 | +0.726 | 0.166 | +87.2 |
| 19 | Ross 578 | -11.49 | +1.458 | -2.697 | 0.062 | -84.5 |
| 20 | Wolf 28 | +05.39 | +1.231 | -2.712 | 0.232 | +263.0 |

**Table 2.** The obtained values of the corresponding parameters for the 20 stars.

| N° | Name | Proper movement ("/year) | Tangential velocity (km/s) | Angle of the total velocity (°) | Total velocity (km/s) |
|---|---|---|---|---|---|
| 1 | Barnard's Star | 10.39 | 90.1 | 140.7 | 142.2 |
| 2 | Kapteyn's Star | 7.32 | 136.6 | 29.1 | 280.6 |
| 3 | Groombridge 1830 | 6.62 | 288.3 | 108.8 | 304.5 |
| 4 | Lacaille 9352 | 5.64 | 87.9 | 83.7 | 88.5 |
| 5 | Ross 619 | 5.20 | 166.9 | 85.0 | 167.6 |
| 6 | 61 CYG A | 4.59 | 76.2 | 130.2 | 99.7 |
| 7 | Lalande 21185 | 4.79 | 57.8 | 145.9 | 103.3 |
| 8 | WX UMA | 3.29 | 76.5 | 47.5 | 103.7 |
| 9 | Wolf 489 | 3.89 | 153.9 | 40.0 | 239.5 |
| 10 | Arcturus | 2.25 | 120.1 | 92.5 | 120.2 |
| 11 | Proxima Centauri | 1.89 | 11.7 | 28.3 | 24.6 |
| 12 | BD +5 1668 | 3.74 | 67.0 | 74.6 | 69.5 |
| 13 | mu CAS | 2.53 | 92.2 | 136.8 | 134.8 |
| 14 | alpha Centauri A | 1.85 | 11.7 | 151.3 | 24.4 |
| 15 | Lacaille 8760 | 2.78 | 52.4 | 68.4 | 56.3 |
| 16 | Luyten 789-6 | 3.20 | 51.7 | 139.2 | 79.1 |
| 17 | Ross 451 | 3.16 | 370.9 | 107.6 | 389.2 |
| 18 | 82 ERI | 2.33 | 66.9 | 37.5 | 109.9 |
| 19 | Ross 578 | 3.05 | 233.7 | 109.9 | 248.5 |
| 20 | Wolf 28 | 2.98 | 60.9 | 13.0 | 269.9 |

As the parallax is inverse to the distance, in Figure 2, we observed that distant objects, with parallax less than 0.3, have total speeds greater than 50 km/s. Furthermore, in Figure 3, we found a good correlation between the tangential speed and the total speed, with a value of 0.73. However, when compared to the modulus of the radial velocity, the value is less than 0.5.

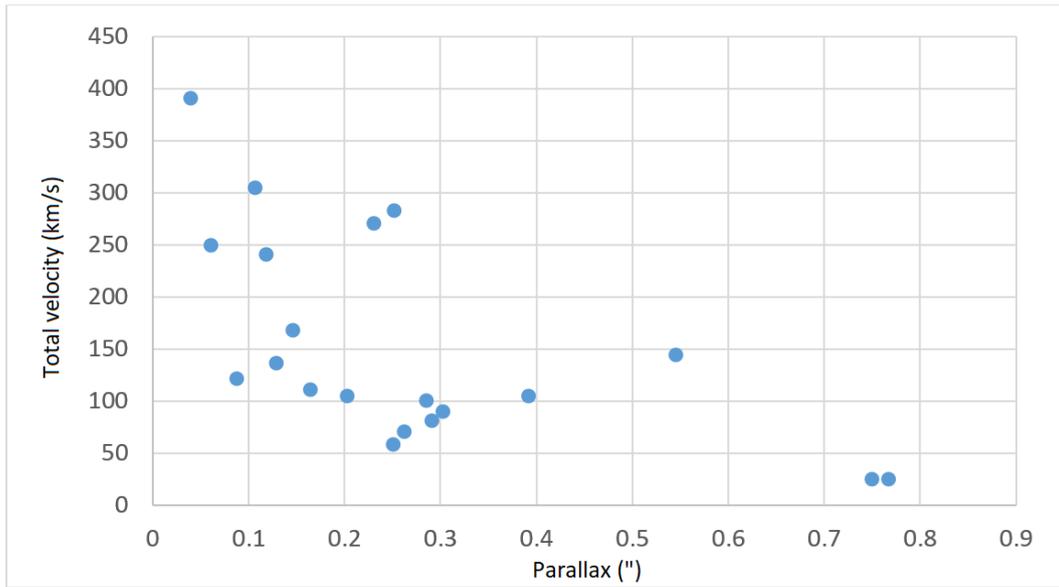

**Figure 2.** Total velocity versus parallax.

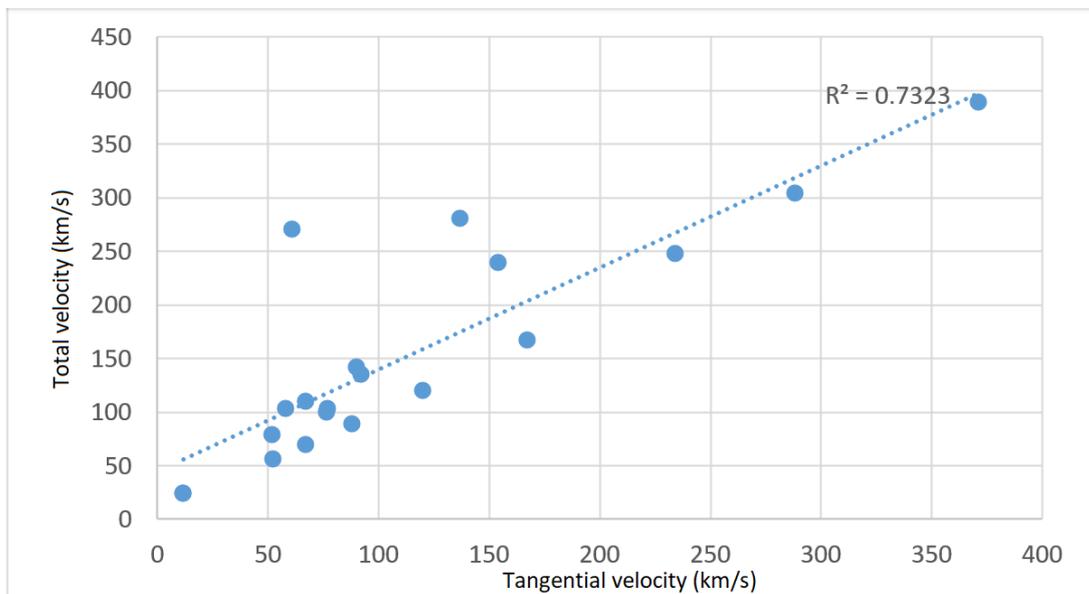

**Figure 3.** Total velocity versus tangential velocity. Symbol $R^2$ stands for the coefficient of determination.

## 5. Conclusions

This research shows that in the context we live in today, where there are new scenarios for learning in the 21st century, what were believed to be great paradigms have ceased to have a certain validity, demonstrating that there are new proposals such as exponential learning.

The present study has shown that the calculation of the relative velocity of a star is possible by using the exponential method. With the algorithm developed in this learning model, the displacement of other forms of learning (paradigms) is showed.

The algorithm introduced is a social algorithm in an open science context as a set of programming rules that allows it to be in the public domain.

In this study, artificial intelligence has not been considered because the problem chosen is not very complex, and it is intended to be an example for a large part of the learning method introduced.

For the data used, we found the more distant stars tend to have a higher relative total velocity, the latter being well correlated with its tangential velocity.

## Acknowledgements

We thank…

# Anexo

## Programming in the Python Language

Next, we present the script in the Python language that automatically solves the problem in astronomy addressed in this paper.

```python
#################################################################################
# Script to calculate the relative speed of a star with respect to the sun #
# by using proper motion data and its radial velocity #
# Dr. Rafael Edgardo Carlos Reyes - - September 2022 #
#################################################################################
# Load the module for working with trigonometric functions in Python
import math

# Define the input parameters:
dec=19.5        # declination
dar =-0.0786    # Variation in right ascension
ddec=-1.996     # Variation in declination
par=0.085       # parallax
vr=-5.1         # radial velocity

# Change the units of time to angular units and calculate the component of proper motion in right ascension
ua=15*(dar)*math.cos(math.pi*dec/180)
print('Proper motion component in right ascension=', ua)

# Define the component of proper motion in declination as the variation in declination
ud=ddec

# Calculate the direction angle of proper motion
theta=180*math.atan(ua/ud)/math.pi
print('Angle of proper motion=', theta)

# By using the signs of the components of proper motion, determine its position angle on the quadrant
         if ua<0 and ud<0:
                 theta=180+theta
                 print('Angle of proper motion=', theta)
         else:
                 print('both are not negative')
# Calculate proper motion
u=ud/math.cos(math.pi*theta/180)
print('Proper motion=', u)

# By using proper motion and parallax, calculate the tangential velocity
vt=4.74*u/par
print('Tangential velocity=', vt)

# Define the components of the velocity of the star
Vsb=vt
Vcb=vr

# Calculate the position angle of the velocity of the star
beta=180*math.atan(Vsb/Vcb)/math.pi

print('Angle of total velocity=', beta)
# Using the velocity components determine its angle of position in the quadrant
         if Vsb>0 and Vcb<0:
                 beta=180+beta
                 print('Angle of total velocity=', beta)
         else:
                 print('It is not in the second quadrant')
```

```python
# Calculation of the total velocity of the star
V=Vsb/math.sin(math.pi*beta/180)
print('Total velocity=', V)
################################################################################
```